\documentclass[12pt]{article}
\usepackage{amsfonts}
\usepackage{amsmath}
\usepackage{amssymb}
\usepackage{myart}

\normalbaselines
\font\tenbm=cmmib10
\font\sevenbm=cmmib7

\newfam\bmfam
\textfont\bmfam=\tenbm \scriptfont\bmfam=\sevenbm
\input tcilatex.tex
\begin{document}

\author{Yuri A. Rylov}
\title{Some subtleties of Riemannian geometry}
\date{Institute for Problems in Mechanics, Russian Academy of Sciences \\
101-1 ,Vernadskii Ave., Moscow, 117526, Russia \\
email: rylov@ipmnet.ru\\
Web site: {$http://rsfq1.physics.sunysb.edu/\symbol{126}rylov/yrylov.htm$}\\
or mirror Web site: {$http://gasdyn-ipm.ipmnet.ru/\symbol{126}%
rylov/yrylov.htm$}}
\maketitle

\begin{abstract}
It is shown, that the conventional presentation of the Maxwell equations for
the electromagnetic field in the Riemannian space-time appears to be
problematic. The reason of hesitations is the fact, that a solution of the
Maxwell equations in the space-time of Minkowski do not turn into solution
of the Maxwell equations in the Riemannian space-time after replacement of
Minkowskian world function $\sigma _{\mathrm{M}}$ by the world function $%
\sigma _{\mathrm{R}}$ of the Riemannian space-time in the solution.
\end{abstract}

\section{Introduction}

The considered problem appeared at an attempt of generalization of dynamics
in the Riemannian space-time on the case of non-Riemannian space-time
geometry. The physical geometry $\mathcal{G}$ is a geometry described in
terms of the world function $\sigma $ and only in terms of $\sigma $ \cite%
{R2001,R2005}. The physical geometry $\mathcal{G}$ and all its relations can
be formulated without a reference to a coordinate system and other means of
description (manifold, dimension, linear vector space). At the conventional
approach to space-time geometries one states that the Riemannian geometry is
the most general type of the space-time geometry. This statement is an
unfounded constraint, because many physical geometries may be considered as
possible space-time geometries. In general, the physical geometry is
multivariant and nonaxiomatizable, whereas the Riemannian geometry pretends
to be an axiomatizable geometry.

The multivariance of a geometry means that at a point $P_{0}$ there are many
vectors $\mathbf{P}_{0}\mathbf{P}_{1},\mathbf{P}_{0}\mathbf{P}_{2},..$ ,
which are equivalent to the vector $\mathbf{Q}_{0}\mathbf{Q}_{1}$ at the
point $Q_{0}$, but vectors $\mathbf{P}_{0}\mathbf{P}_{1},\mathbf{P}_{0}%
\mathbf{P}_{2},..$are not equivalent between themselves. Vectors are
equivalent ($\mathbf{P}_{0}\mathbf{P}_{1}$eqv$\mathbf{Q}_{0}\mathbf{Q}_{1}$%
), if vectors $\mathbf{P}_{0}\mathbf{P}_{1}$ and $\mathbf{Q}_{0}\mathbf{Q}%
_{1}$ are parallel and their lengths $\left\vert \mathbf{P}_{0}\mathbf{P}%
_{1}\right\vert $ and $\left\vert \mathbf{Q}_{0}\mathbf{Q}_{1}\right\vert $
are equal%
\begin{equation}
\mathbf{P}_{0}\mathbf{P}_{1}\uparrow \uparrow \mathbf{Q}_{0}\mathbf{Q}%
_{1}:\qquad (\mathbf{P}_{0}\mathbf{P}_{1}.\mathbf{Q}_{0}\mathbf{Q}%
_{1})=\left\vert \mathbf{P}_{0}\mathbf{P}_{1}\right\vert \cdot \left\vert 
\mathbf{Q}_{0}\mathbf{Q}_{1}\right\vert  \label{a1.1}
\end{equation}%
\begin{equation}
\left\vert \mathbf{P}_{0}\mathbf{P}_{1}\right\vert =\left\vert \mathbf{Q}_{0}%
\mathbf{Q}_{1}\right\vert  \label{a1.2}
\end{equation}%
where the scalar product $(\mathbf{P}_{0}\mathbf{P}_{1}.\mathbf{Q}_{0}%
\mathbf{Q}_{1})$ and the length $\left\vert \mathbf{P}_{0}\mathbf{P}%
_{1}\right\vert $ of the vector $\mathbf{P}_{0}\mathbf{P}_{1}$ are defined
by the relations 
\begin{equation}
(\mathbf{P}_{0}\mathbf{P}_{1}.\mathbf{Q}_{0}\mathbf{Q}_{1})=\sigma \left(
P_{0},Q_{1}\right) +\sigma \left( P_{1},Q_{0}\right) -\sigma \left(
P_{0},Q_{0}\right) -\sigma \left( P_{1},Q_{1}\right)  \label{a1.3}
\end{equation}%
\begin{equation}
\left\vert \mathbf{P}_{0}\mathbf{P}_{1}\right\vert ^{2}=(\mathbf{P}_{0}%
\mathbf{P}_{1}.\mathbf{P}_{0}\mathbf{P}_{1})=2\sigma \left(
P_{0},P_{1}\right)  \label{a1.4}
\end{equation}%
Let us stress, that the equivalence (\ref{a1.1}), (\ref{a1.2}) of two
vectors is defined in terms of the world function $\sigma $ and only in
terms $\sigma $, which is defined as follows 
\begin{equation}
\sigma :\qquad \Omega \times \Omega \rightarrow \mathbb{R},\qquad \sigma
\left( P,P\right) =0,\qquad \forall P\in \Omega  \label{a1.5}
\end{equation}%
Here $\Omega $ is the set of points, where the physical geometry $\mathcal{G}
$ is given. The world function is interpreted in the form $\sigma \left(
P,Q\right) =\frac{1}{2}\rho ^{2}\left( P,Q\right) $, where $\rho \left(
P,Q\right) $ is the distance between the points $P$ and $Q$.

In the proper Euclidean geometry the equivalence relation (\ref{a1.1}), (\ref%
{a1.2}) coincides with the conventional definition of two vector
equivalence, which is formulated as equality of the vector components in the
Cartesian coordinate system%
\begin{equation}
\mathbf{p}=\mathbf{q,\ }\text{if\qquad }p_{i}=q_{i},\qquad i=1,2,...n
\label{a1.6}
\end{equation}%
where $p_{i}$ and $q_{i}$ are coordinates of vectors $\mathbf{p}$ and $%
\mathbf{q}$ in some Cartesian coordinate system, and $n$ is the dimension of
the proper Euclidean geometry.

The definition (\ref{a1.1}), (\ref{a1.2}) distinguishes from the
conventional definition (\ref{a1.6}) in the relation, that the definition (%
\ref{a1.1}), (\ref{a1.2}) does not contain such auxiliary means of
description as dimension, coordinate system and concept of the linear vector
space. Besides, the definition (\ref{a1.1}), (\ref{a1.2}) contains two
equations for the proper Euclidean geometry of any dimension, whereas in the
conventional definition the number of equations depends on the dimension of
the space. All this means that the definition (\ref{a1.1}), (\ref{a1.2}) is
more fundamental, than the definition (\ref{a1.6}), which can be used, only
if in the geometry one can introduce concept of the linear vector space with
the scalar product, given on it. The relation of equivalence (\ref{a1.1}), (%
\ref{a1.2}) can be used in any physical geometry, whereas the equivalence
relation (\ref{a1.6}) can be used only in the space-time geometry, where the
linear vector space can be introduced.

In general, the physical geometry is multivariant, because the definition (%
\ref{a1.1}), (\ref{a1.2}) admits an existence of many vectors $\mathbf{P}_{0}%
\mathbf{P}_{1}$, which are equivalent to the given vector $\mathbf{Q}_{0}%
\mathbf{Q}_{1}$. If the physical geometry $\mathcal{G}$ is multivariant, the
equivalence relation (\ref{a1.1}), (\ref{a1.2}) is intransitive. In this
case the physical geometry $\mathcal{G}$ cannot be axiomatizable, because in
any axiomatizable geometry the equivalence relation is transitive.

In the Riemannian geometry the world function $\sigma _{\mathrm{R}}$ is
defined by the relation%
\begin{equation}
\sigma _{\mathrm{R}}\left( P,Q\right) =\frac{1}{2}\left( \dint\limits_{%
\mathcal{L}_{\left[ PQ\right] }}\sqrt{g_{ik}dx^{i}dx^{k}}\right) ^{2}
\label{a1.7}
\end{equation}%
where $\mathcal{L}_{\left[ PQ\right] }$ is a segment of the geodesic,
connecting the point $P$ and $Q$. One can construct the Riemannian geometry
as a physical geometry $\mathcal{G}_{\sigma \mathrm{R}}$, using the relation
(\ref{a1.7}), as a definition of the world function. We shall refer to the
geometry $\mathcal{G}_{\sigma \mathrm{R}}$ as the $\sigma $-Riemannian
geometry, which distinguishes from the conventional construction of the
Riemannian geometry.

The conventional Riemannian geometry is constructed as a set of
infinitesimal Euclidean geometries, "glued" between themselves in some
manner. The manner of conglutination determines the peculiar properties of
the Riemannian geometry. This manner of conglutination is described by the
character of the dependence of the metric tensor on the coordinates. In
general, the Riemannian geometry appears to be multivariant in the sense,
that equivalence of remote vectors depends on the path of their parallel
transport. To remove multivariance of the Riemannian geometry, the
equivalence relation of the remote vectors is removed. As a result the
conventional Riemannian geometry pretends to be single-variant and
axiomatizable. However, such an approach is not consecutive, because the
multivariant geometry is nonaxiomatizable, and one cannot turn the
nonaxiomatizable geometry into axiomatizable one by the prohibition of the
remote vector equivalence.

Thus, the $\sigma $-Riemannian geometry is multivariant, in general, and
consistent. The $\sigma $-Riemannian geometry cannot be inconsistent in
principle, because it is not deduced from axiomatics. Inconsistency of a
geometry is an attribute of the geometry construction method, when the
geometry is deduced from a system of axiom. The $\sigma $-Riemannian
geometry is constructed as a deformation of the proper Euclidean geometry.
All propositions $\mathcal{P}_{\mathrm{E}}$ of the proper Euclidean geometry 
$\mathcal{G}_{\mathrm{E}}$ are presented in terms of the Euclidean world
function $\sigma _{\mathrm{E}}$ in the form $\mathcal{P}_{\mathrm{E}}\left(
\sigma _{\mathrm{E}}\right) $. Replacing $\sigma _{\mathrm{E}}\ $by the
world function $\sigma _{\sigma \mathrm{R}}$ of the $\sigma $-Riemannian
geometry, one obtains all propositions $\mathcal{P}_{\mathrm{E}}\left(
\sigma _{\sigma \mathrm{R}}\right) $ of the $\sigma $-Riemannian geometry.
Procedure of deformation does not use logical reasonings, and it cannot be
inconsistent in principle.

The conventional Riemannian geometry is single-variant, but inconsistent.
This inconsistency manifests itself, in particular, in the problem of
generalization of dynamics in the Riemannian space-time on the case of
arbitrary space-time physical geometry.

\section{Generalization of dynamics in the Riemannian space-time on the case
of arbitrary space-time}

If the space-time geometry may be an arbitrary physical geometry, we are to
generalize the dynamics in the Riemannian space-time on the case of
arbitrary physical space-time geometry. The first part of this
generalization (motion of a pointlike particle in the given classical
fields: gravitational and electromagnetic) was made successfully in \cite%
{R2008}. This generalization leads to the difference dynamic equations. It
is quite reasonable, because the space-time geometry may be discrete, and
differential dynamic equations in the discrete space-time geometry are not
natural, whereas the difference dynamic equations are suitable in both
continuous and discrete space-time geometry. Such a generalization gives
rather unexpected results. It appears, that the proper choice of the
space-time geometry, which is free of unfounded constraints of the
Riemannian geometry, admits one to explain quantum effects as a statistical
description of multivariant motion of particles, generated by the
multivariance of the space-time geometry. Besides, arrangement of elementary
particle is determined by the structure of its skeleton. The skeleton is a
set of several points in the space-time. The mutual displacement of these
points determines structure of the skeleton \cite{R2008}. The quantum
properties (wave function, quantization, renormalization) appear to be
needless. In particular, the Dirac particle is composite \cite{R2008a}. Its
skeleton consists of three points. World chain of such a particle is a
spacelike helix with the timelike axis. Thus, generalization, suggested in 
\cite{R2008}, realize the generalization of the special relativity on the
case of the arbitrary physical space-time geometry.

The second part of the generalization is a consideration of the influence of
the matter distribution on the space-time geometry. The general relativity
considers this influence in the framework of the Riemannian space-time
geometry. One needs to generalize the general relativity on the case of the
arbitrary space-time geometry. In principle this problem is solved by
representation of the Maxwell equations and the gravitation equations in
terms of the world function $\sigma _{\sigma \mathrm{R}}$ of the Riemannian
space-time geometry. Thereafter the world function $\sigma _{\sigma \mathrm{R%
}}$ is replaced by the world function $\sigma $ of arbitrary physical
space-time geometry. Then one needs to choose such a space-time geometry,
which agrees with the experimental data.

However, an attempt of generalization of Maxwell equations for the
electromagnetic field meets a difficulty. In the space-time geometry of
Minkowski the dynamic equations for the electromagnetic potential $A_{k}$
have the form 
\begin{equation}
g_{\mathrm{M}}^{ik}\partial _{i}\partial _{k}A_{l}=\frac{4\pi }{c}%
j_{l},\qquad F_{ik}=\partial _{i}A_{k}-\partial _{k}A_{i}  \label{a2.1}
\end{equation}%
where $F_{ik}$ is the tensor of the electromagnetic field, and $j_{l}\left(
x\right) $ is the 4-vector of the electric current, generating the
electromagnetic field. The world function between the points $x$ and $%
x^{\prime }$ has the form%
\begin{equation}
\sigma _{\mathrm{M}}\left( x,x^{\prime }\right) =\frac{1}{2}g_{\mathrm{M}%
ik}\left( x^{i}-x^{\prime i}\right) \left( x^{k}-x^{\prime k}\right) 
\label{a2.2}
\end{equation}%
The first equation (\ref{a2.1}) can be integrated in the form%
\begin{equation}
A_{l}\left( x\right) =-\frac{4\pi }{c}\dint G_{\mathrm{ret}}\left(
x-x^{\prime }\right) j_{l}\left( x^{\prime }\right) d^{4}x^{\prime }
\label{a2.3}
\end{equation}%
where the retarded Green function $G_{\mathrm{ret}}\left( x-x^{\prime
}\right) $ satisfies the equation 
\begin{equation}
g_{\mathrm{M}}^{ik}\partial _{i}\partial _{k}G_{\mathrm{ret}}\left(
x-x^{\prime }\right) =-\delta ^{\left( 4\right) }\left( x-x^{\prime }\right)
=-\dprod\limits_{i=0}^{i=3}\delta \left( x^{i}-x^{\prime i}\right) 
\label{a2.4}
\end{equation}%
and has the form 
\begin{equation}
G_{\mathrm{ret}}\left( x-x^{\prime }\right) =\frac{1}{2\pi }\theta \left(
x^{0}-x^{0\prime }\right) \delta \left( 2\sigma _{\mathrm{M}}\left(
x,x^{\prime }\right) \right)   \label{a2.5}
\end{equation}%
\begin{equation*}
\theta \left( x\right) =\left\{ 
\begin{array}{ccc}
1 & \text{if} & x>0 \\ 
0 & \text{if} & x\leq 0%
\end{array}%
\right. 
\end{equation*}%
Here for simplicity we consider the case, when the nonvanishing 4-current
density is concentrated inside small spatial region.

To write dynamic equations (\ref{a2.1}) in the Riemannian space-time
geometry, conventionally one replaces the partial derivatives by covariant
derivatives. One obtains instead of (\ref{a2.1})%
\begin{equation}
g^{ik}\nabla _{i}\nabla _{k}A_{l}=\frac{4\pi }{c}j_{l},\qquad F_{ik}=\nabla
_{i}A_{k}-\nabla _{k}A_{i}=\partial _{i}A_{k}-\partial _{k}A_{i}
\label{a4.6}
\end{equation}%
where $\nabla _{k}$ means the covariant derivative.

It is rather difficult to express differential equations in terms of the
world function $\sigma _{\sigma \mathrm{R}}=\sigma _{\mathrm{R}}$ of the $%
\sigma $-Riemannian geometry, because the world function is an integral
(two-point) quantity. The finite relations and integral relations are
expressed in terms of the world function more effective. Let us replace the
world function $\sigma _{\mathrm{M}}$ by the world function $\sigma _{%
\mathrm{R}}$ in the expression (\ref{a2.5}) and substitute the obtained
expression in the relation of the type of (\ref{a2.4}). We omit the first
factor $\frac{1}{2\pi }\theta \left( x^{0}-x^{0\prime }\right) $, because it
gives a contribution to dynamic equation only at $x^{0}=x^{0\prime }$. We
obtain 
\begin{eqnarray}
&&g^{ik}\nabla _{i}\nabla _{k}\left( \delta \left( 2\sigma _{\mathrm{R}%
}\left( x,x^{\prime }\right) \right) \right)  \notag \\
&=&g^{ik}\nabla _{i}\left( \delta ^{\prime }\left( 2\sigma _{\mathrm{R}%
}\left( x,x^{\prime }\right) \right) 2\sigma _{\mathrm{R}|k}\right)  \notag
\\
&=&g^{ik}\left( \delta ^{\prime \prime }\left( 2\sigma _{\mathrm{R}}\left(
x,x^{\prime }\right) \right) 4\sigma _{\mathrm{R}|k}\sigma _{\mathrm{R}%
|i}+\delta ^{\prime }\left( 2\sigma _{\mathrm{R}}\left( x,x^{\prime }\right)
\right) 2\sigma _{\mathrm{R}|k|i}\right)  \label{a2.7}
\end{eqnarray}%
where the vertical stroke means the covariant derivative. 
\begin{equation*}
\sigma _{\mathrm{R}|k}=\nabla _{k}\sigma _{\mathrm{R}}\left( x,x^{\prime
}\right) =\frac{\partial \sigma _{\mathrm{R}}\left( x,x^{\prime }\right) }{%
\partial x^{k}}
\end{equation*}%
We use the world function $\sigma _{\mathrm{R}}$ instead $\sigma _{\sigma 
\mathrm{R}}$, because these quantities coincide.

Let us take into account the identity 
\begin{equation}
x\delta ^{\prime \prime }\left( x\right) +2\delta ^{\prime }\left( x\right)
=0  \label{a4.10}
\end{equation}%
and the fact, that the Riemannian world function satisfies the relation \cite%
{S60}%
\begin{equation}
\sigma _{\mathrm{R}|k}g^{ik}\sigma _{\mathrm{R}|i}=2\sigma _{\mathrm{R}}
\label{a4.9}
\end{equation}
We obtain from (\ref{a2.7})%
\begin{equation}
g^{ik}\nabla _{i}\nabla _{k}\left( \delta \left( 2\sigma _{\mathrm{R}}\left(
x,x^{\prime }\right) \right) \right) =2\delta ^{\prime }\left( 2\sigma _{%
\mathrm{R}}\left( x,x^{\prime }\right) \right) \left( g^{ik}\sigma _{\mathrm{%
R}|k|i}-4\right)  \label{a4.11}
\end{equation}%
If the Riemannian space-time coincides with the 4-dimensional space-time of
Minkowski, the rhs of (\ref{a4.11}) vanishes, because the quantity $%
g^{ik}\sigma _{\mathrm{M}|k|i}$ is a scalar, which in the inertial
coordinate system has the form 
\begin{equation}
g^{ik}\sigma _{\mathrm{M}|k|i}=4  \label{a4.12}
\end{equation}%
Note, that the relation (\ref{a4.12}) takes place only in the 4-dimensional
space-time of Minkowski.

In the case of arbitrary Riemannian space-time the equation is not valid, in
general, and rhs of (\ref{a4.11}) does not vanish, in general. It means,
that the transition from the space-time of Minkowski to the Riemannian
space-time by means of replacement of the world function of Minkowski in the
relations (\ref{a2.3}) - (\ref{a2.5}) and the replacement procedure of
partial derivatives by the covariant ones in the dynamic equations (\ref%
{a2.1}) are different procedures.

Thus, writing the Maxwell equations in terms of the world function, we are
to choose between two alternatives: (1) use of conventional representation
of the Riemannian geometry, which pretends to a single--variance, which is
not single-variant in reality and (2) use of $\sigma $-Riemannian geometry,
which multivariant and nonaxiomatizable, in general.

The $\sigma $-Riemannian geometry has the advantage of the Riemannian
geometry in the sense, that it is consistent, whereas the Riemannian
geometry is inconsistent.

Procedure of deduction of dynamic equations (\ref{a4.6}) in the Riemannian
geometry is founded on a use of curvilinear coordinate system. Dynamic
equations (\ref{a2.1}) are written in the curvilinear coordinate system of
the space-time of Minkowski in the form (\ref{a4.6}). Thereafter one
declares, that the form (\ref{a4.6}) of dynamic equations is valid in
arbitrary Riemannian space-time. However, it appears, that this declaration
is incompatible with the replacement of the world function $\sigma _{\mathrm{%
M}}$ by the world function $\sigma _{\mathrm{R}}$ in solutions of dynamic
equations (\ref{a4.6}) in the Minkowski space-time. A use of the coordinate
system in deduction of dynamic equations (\ref{a4.6}) in the case of the
Riemannian geometry seems to be problematic (compare the role of coordinate
system (\ref{a1.6}) in the definition of equivalence (\ref{a1.1}), (\ref%
{a1.2})).

\section{Concluding remarks}

Thus, trying to generalize the Maxwell equation on the case of
non-Riemannian geometry, we meet unexpected problem, that the conventional
presentation of the Maxwell equations in the Riemannian geometry appears to
be problematic. It is possible, that, the same problem will appear at an
attempt of generalization of the gravitation equation. One should look for
the way around these problems.

\end{document}